# Mott Gap Softening Coinciding with Spin Correlations Collapse in α-$RuCl_3$

Wei Yue[1], Xiaohu Zheng[2,4*], Chongli Yang[2], Kun Peng[1], Rui-Rui Du[1,3*]

[1]International Center for Quantum Materials, School of Physics, Peking University, Beijing 100871, China.

[2]Beijing Academy of Quantum Information Sciences, Beijing 100193, China.

[3]CAS Center for Excellence in Topological Quantum Computation, University of Chinese Academy of Sciences, Beijing 100190, China.

[4]State Key Laboratory of Materials for Integrated Circuits，Shanghai Institute of Microsystem and Information Technology，Chinese Academy of Sciences，865 Changning Road, Shanghai 200050 China

[*]Corresponding authors: zhengxh@baqis.ac.cn; rrd@pku.edu.cn.

**The Mott insulator α-$RuCl_3$, featuring the intertwined interplay of spin-orbit coupling (SOC) and Kitaev spin correlations, provides an unparalleled platform for probing quantum many-body physics. Using scanning tunneling microscopy/spectroscopy (STM/STS), we compare temperature-dependent *dI/dV* spectra between *in-situ* grown monolayers and exfoliated bulk samples. Both systems exhibit pronounced Mott gap softening near 110 K, manifested by spectral weight transfer from Hubbard bands toward the Fermi level, resulting in low-energy correlated charge delocalization. Although this gap softening coincides with Kitaev paramagnetic and structural phase transitions in bulk crystals, monolayer studies provide compelling insights. By eliminating structural phase transition in monolayer sample, we suggest that spin correlations, rather than Coulomb interactions alone, may govern charge dynamics within the Mott-**

**Hubbard framework, challenging conventional Mott-Hubbard paradigms. These results resolve a long-standing controversy regarding the Mott gap magnitude in α-$RuCl_3$ and experimentally confirm the critical role of spin correlations in Mott physics.**

## I. INTRODUCTION

Mott insulators and their metal-to-insulator transitions (MIT) constitute intricate phenomena driven by strong electron correlations, representing both a central research focus and a fundamental challenge in condensed matter physics [1–3]. Mott's seminal work attributed the insulating state primarily to strong on-site Coulomb repulsion in half-filled electron bands, inducing charge localization and defining the quintessential Mott insulator [4]. In contrast, theorists like Slater proposed that magnetic ordering alone could generate charge gaps through perturbative treatment of Coulomb interactions [5]. The Mott physics are further enriched by spin–orbit coupling (SOC) in low-dimensional systems, where quantum confinement can facilitate unique spin–charge separation manifested by emergent fractionalized quasiparticles—spinons and chargons [6]. These excitations exhibit distinct dispersion characteristics while maintaining intrinsic connections to their parent electrons. Such fractionalization enables novel quantum phenomena, including quantum spin liquid (QSL) states [7–11] and topological superconductivity [12–15], motivating re-examination of Mott physics through the dual perspective of charge, spin, and their intercorrelations [16–21], thereby extending beyond traditional Mott and Slater paradigms.

As a leading Kitaev QSL candidate emerging from the SOC assisted Mott parent state, α-$RuCl_3$ has garnered significant research interest [22–26]. Substantial progress in characterizing its magnetic properties has established a clear phase diagram (Fig. 1a), featuring three distinct regimes: (i) Below 80–120 K, short-range Ising spin correlations dominate, generating strong spin frustration and bringing the system into close proximity to the Kitaev QSL phase-denoted as the Kitaev quantum paramagnetic (KQPM) state [7,27–30]; (ii) Above this temperature range, spin correlations become negligible upon the emergence of a conventional paramagnetic (CPM) state; (iii) Long-range antiferromagnetic order emerges below 7 K due to increasing non-Kitaev interactions [28,29,31]. Therefore, α-$RuCl_3$ provides a unique magnetic Mott insulator system exhibiting transitions from an antiferromagnet through the KQPM phase to a CPM phase at easily accessible temperatures (below room temperature). Additionally, α-$RuCl_3$ is a 4*d* transition-metal halide insulator featuring honeycomb layers of edge-sharing $RuCl_6$ octahedra [26], providing an ideal platform for studying intrinsic electronic properties and many-body correlations in the two-dimensional limit (as shown in Fig. 1b). These distinctive characteristics establish α-$RuCl_3$ as a premier system for exploring Mott physics through the unique charge-spin correlations. However, fundamental aspects of its electronic structure remain unresolved, particularly regarding the Mott gap, where reported values span 0.2–2.2 eV without consensus [24–26,32–37].

In this study, we employ STM/STS to investigate both transferred bulk and *in-situ* grown monolayer α-$RuCl_3$ samples. Our measurements reveal a strong temperature dependence of the Mott gap, characterized by rapid softening near 110 K. Notably, negligible gap changes occur near 7 K, despite magnetic phase transitions involving both antiferromagnetic and KQPM states. This pronounced softening leads to a dramatic reduction in the gap magnitude, from approximately 2 eV at 4.2 K to nearly gapless at room temperature. Since the monolayer α-$RuCl_3$ deposited on pristine graphite effectively circumvents interference from structural phase transitions inherent in bulk crystals, the cause of the band gap softening can be attributed to the loss of spin correlations. Temperature-dependent *dI/dV* spectra indicate that the gap softening arises from spectral weight transfer from Hubbard bands toward the Fermi level, while preserving the fundamental Mott framework. This spectral shift weakens charge localization within the Mott-Hubbard picture. Furthermore, the direct correlation between gap softening and the loss of spin correlations establishes that both long-range ordered and short-range frustrated spin correlations in the framework of SOC Mott insualtors may critically impact the gap formation. Our findings not only reconcile significant discrepancies in reported α-$RuCl_3$ energy gaps but also confirm the crucial role of spin correlations in governing Mott physics. These results motivate further exploration of emergent phenomena arising from spin-charge entanglement in Mott insulators.

## II. *dI/dV* SPECTRA ON TRANSFERRED THICK α-$RuCl_3$ FLAKES

Transferred α-$RuCl_3$ samples were prepared following established protocols detailed in our previous work [37–39]. To access the intrinsic bulk electronic properties while minimizing hybridization effects from the graphite substrate, STM/STS investigations were conducted on these transferred thick flakes (>5 nm). The *dI/dV* spectra in Fig. 1c exhibit a pronounced temperature dependence of the Mott gap, evolving strikingly from a well-defined hard gap (~2 eV) at low temperatures to a nearly gapless state at elevated temperatures (e.g. 300 K). This unconventional evolution cannot be attributed to conventional thermal fluctuation effects or the magnetic-order-driven Slater mechanism [5], as negligible gap changes are observed across the magnetic phase transition (from antiferromagnet to KQPM state at ~7 K) within the temperature range of 4.2 K to 77 K. Critically, this strong temperature dependence reconciles significant discrepancies in previously reported Mott gap values [24–26,30–35]. Specifically, the low-temperature results exhibit a comparably large gap to those observed in photoemission measurements [34,36]; while the room-temperature *dI/dV* spectrum in the current work aligns well with the previous *dI/dV* spectrum reported in Ref. [40]. Continuous temperature-dependent STS measurements (Fig. 1d) further demonstrate that the gap softening occurs rapidly within a narrow temperature window from approximately 95 to 110 K. This temperature range coincides precisely with the magnetic transition (from KQPM state to CPM state, Fig. 1a) [29,41–47], but interferenced by the structural phase transition (from *C2/m* at high temperature to $R\bar{3}$ at low temperature, as shown in Fig. 1b) observed in bulk crystals [22,48–51]. Consequently, decoupling the respective contributions of structural and magnetic phase transitions presents a fundamental challenge essential for elucidating the mechanism of temperature-dependence of the Mott gap in α-$RuCl_3$. Importantly, the structural phase transition is only closely tied to the bulk crystal, while the magnetic phase transition remains unaffected by decreasing layer thickness—the Kitaev interaction in monolayer can even be enhanced [52].

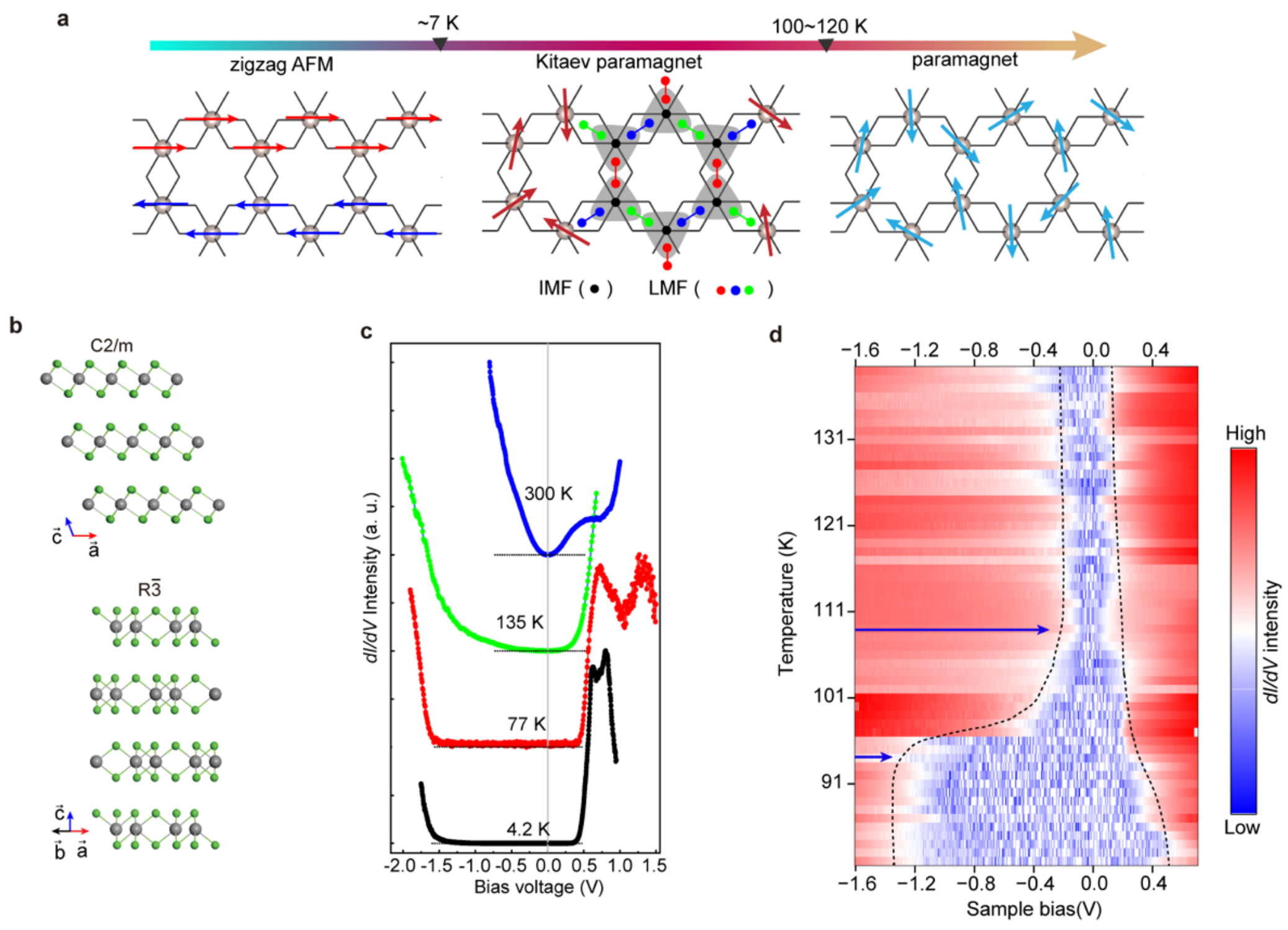

Fig. 1. Temperature-dependent STS on transferred α-$RuCl_3$ flakes. (a) Phase diagram of the spin dynamics in α-$RuCl_3$ as a function of temperature, at temperature lower than 7 K, the compound becomes a antiferromagnet with zigzag spin orders; at temperature higher than approximately 110 K, the compound goes into the CPM phase; in the medium temperature between 7 to 110 K, it is a KQPM state with short-ranged Ising spin coupling; (b) Layer stacking of bulk form α-$RuCl_3$ in the monoclinic ($C2/m$) crystal structure at high-temperature and in therhombohedral ($R\bar{3}$) crystal structure at low temperatures; (c) *dI/dV* spectra taken on thick transferred α-$RuCl_3$ flake at different temperatures; (d) *dI/dV* spectra as the function of temperatures taken on transferred thick α-$RuCl_3$ ($V_{bias}$ = 0.6 V, $I_{set}$ = 600 pA) .

## III. SYNTHESIZATION OF MONOLAYER α-$RuCl_3$

A monolayer of α-$RuCl_3$ was then synthesized on a cleaved highly oriented pyrolytic graphite (HOPG) substrate *via* thermal evaporation of purified anhydrous α-$RuCl_3$ powder at 350 °C in a K-cell under a base chamber vacuum of $6\times10^{-10}$ Torr. The substrate temperature was maintained at 220 °C during deposition, followed by in-situ annealing at the same temperature for 45 minutes. The sample was then transferred to a STM chamber and cooled to the liquid nitrogen (77 K) and liquid helium temperatures (4.2 K) for STM/STS measurements. The *dI/dV* spectra were acquired using lock-in amplification with a modulation frequency of 707 Hz and an amplitude modulation of 5-10 mV. Temperature-dependent *dI/dV* spectra were collected during controlled warming, ensuring thermal stability at each measuring point.

The large-scale STM image (Fig. 2a) reveals α-$RuCl_3$ thin films on graphite with characteristic trapezoidal voids. These films comprise multiple crystalline domains, as indicated by color-coded dashed lines tracing the void edges in Fig. 2a. Effective suppression of β-$RuCl_3$ impurities was achieved through pre-purification of commercial α-$RuCl_3$ powder. The height profile in Fig. 2c, measured along the white dashed line in Fig. 2b, confirms a uniform thickness of ~800 pm consistent with monolayer α-$RuCl_3$. High-resolution STM (Fig. 2d) shows vacancy defects surrounded by blurred quasiparticle interference patterns, while the fast Fourier transform (FFT) inset preserves hexagonal symmetry indicating the absence of crystal reconstruction. Atomic-resolution imaging (Fig. 2e) resolves a honeycomb lattice of Ru atoms, demonstrating crystallographic equivalence to cleaved bulk samples. Notably, the trapezoidal void edges align with the armchair directions of this honeycomb lattice.

We acquired *dI/dV* spectra at three characteristic temperatures (Fig. 2f). A significant spectral contrast in gap magnitude is observed: approximately 2 eV for bulk (Fig. 1b) [37] versus 0.6 eV for the monolayer at low temperature. This contrast originates from graphite/α-$RuCl_3$ heterointerface coupling in the monolayer system [38,53–55]. Both the 110 K and 4.2 K spectra exhibit comparable hard gaps exceeding 0.6 eV, again ruling out the thermal fluctuations as the predomiant origin of the gap variation. Furthermore, the elimination of long-range spin order at approximately 7 K in the monolayer sample didn't alter the *dI/dV* spectra. In contrast, the 159 K spectrum exhibits pronounced gap softening. Crucially, comparison of atomic-resolution STM images at 300 K and 4.2 K (insets in Fig. 2f) confirms the absence of serious temperature-induced lattice reconstruction. These results collectively rule out both single-layer lattice reconstruction and bulk structural phase transitions as potential origins of the gap softening, thereby establishing this gap softening as an intrinsic response of α-$RuCl_3$.

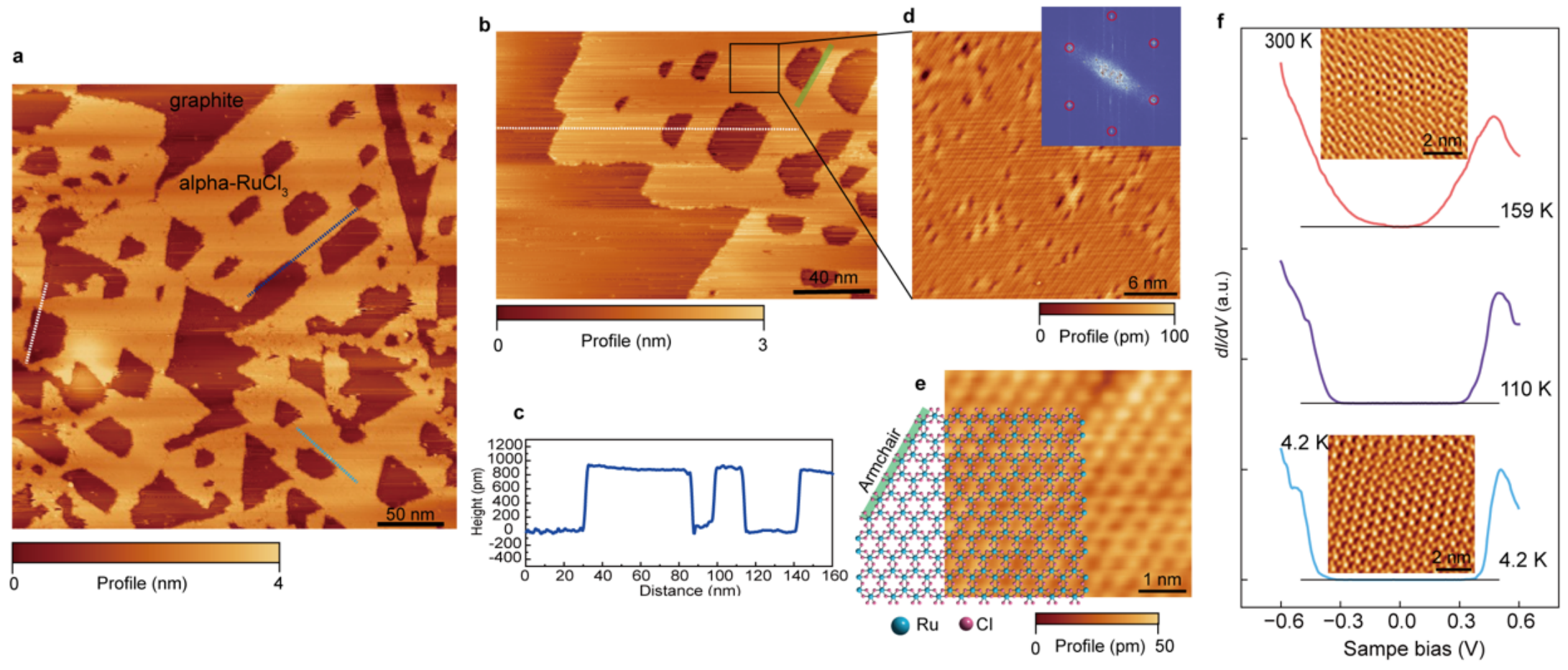

Fig. 2. Monolayer α-$RuCl_3$ *in-situ* grown on graphite surface. (a) Large scale STM

morphology of the growth of α-$RuCl_3$ on graphite substrate ($V_{bias}$ = 1.5 V, $I_{set}$ = 200 pA). The colored dashed lines along the edges of trapezoidal interspaces indicate the lattice orientations of different crystal domains; (b) An enlarged STM image of the α-$RuCl_3$ domain on graphite ($V_{bias}$ = 1.5 V, $I_{set}$ = 200 pA); (c) Height profile along the white dashed line in (b) illustrates the thickness of the α-$RuCl_3$ film; (d) Enlarged STM image of the square area in (b) ($V_{bias}$ = 1.5 V, $I_{set}$ = 100 pA); inset shows a FFT of the image; (e) STM atomic-resolved morphology of the α-$RuCl_3$ ($V_{bias}$ = 1.5 V, $I_{set}$ = 100 pA). The lattice constructions are presented with the indication of armchair edge; (f) Normalized *dI/dV* spectra taken on monolayer α-$RuCl_3$ at three different temperatures(4.2 K: $V_{bias}$ = 0.75 V, $I_{set}$ = 500 pA; 110 K, 159 K: $V_{bias}$ = 0.5 V, $I_{set}$ = 300 pA); insets show atomic-resolution STM images at 300 K and 4.2 K.

## IV. TEMPERATURE-DEPENDENT *dI/dV* SPECTRA ON MONOLAYER α-$RuCl_3$

To precisely monitor the onset temperature and evolution of gap softening in monolayer α-$RuCl_3$, we systematically acquired continuous temperature-dependent *dI/dV* spectra. A series of spectra were collected in the temperature range spanning the transition from KQPM state to CPM state wherein the gap solftening occurs. Temperature-dependent spectral mapping (Fig. 3a) reveals a similar abrupt softening of the Mott gap in the monolayer α-$RuCl_3$, closely resembling that the bulk sample and occurring precisely near the Kitaev magnetic transition temperature ($T_k$ ~110 K) [56,57]. Compared to bulk samples, this softening process occurs more abruptly and at a silightly higher transition temperature. We attribute these differences to the elimination of interlayer hopping and the potential enhancement of Kitaev interactions in the monolayer [52,56]. Notably, a remaining hard gap persists in the monolayer even at room temperature (Fig. 3c), unlike in bulk samples. We propose that three dimensional interlayer hopping of delocalized electrons in bulk crystals has further softened the charge gap. Crucially, two distinct electronic regimes emerge in Fig. 3a: a stabilized hard gap phase below $T_k$ and a renormalized soft gap phase above $T_k$ with minimal thermal hysteresis within each phase. Quantitative analysis of the Mott gap temperature dependence reveals striking correspondence with the anomalous resistance transition in Ref. [58] as shown in Fig. 3b, where the resistance plunge directly tracks gap softening. This demonstrates that the renormalized soft gap governs charge transport beyond conventional band transport paradigms.

Fig. 3c demonstrates that gap softening generates low-energy in-gap states while preserving the energy positions of the gap-edge states—specifically the upper Hubbard band (UHB) and lower Hubbard band (LHB). This observation provides a reasonable explanation for the lack of significant temperature dependence in Mott-Hubbard bands observed in photoelectron spectroscopy techniques [32,34,36,59], which contrasts with the elimination of hard gap observed in STS [40,60]. The spectral rigidity of Hubbard bands is further resolved in the logarithmic-scale plots, as shown in Fig. 3d. A close-up view of the UHB (shaded region, Fig. 3c), presented in Fig. 3e, reveals significant spectral weight transfer toward the Fermi level during gap softening, with minimal change on the high-energy flank. The spectral weight transfer from HBs to the Fermi level characterizes the gap softening as a many-body renormalization process, wherein localized band-edge states become weakly delocalized into the gap, driving coherent charge redistribution. Owing to the strong correlation in the Mott-Hubbard framework, this weakening of charge localization can enhances conductivity while maintaining ultralow carrier mobility ($\mu < 1$ $cm^2V^{-1}s^{-1}$) and an anomalously high effective electron mass, as inferred from transport experiments in refs. [61,62].

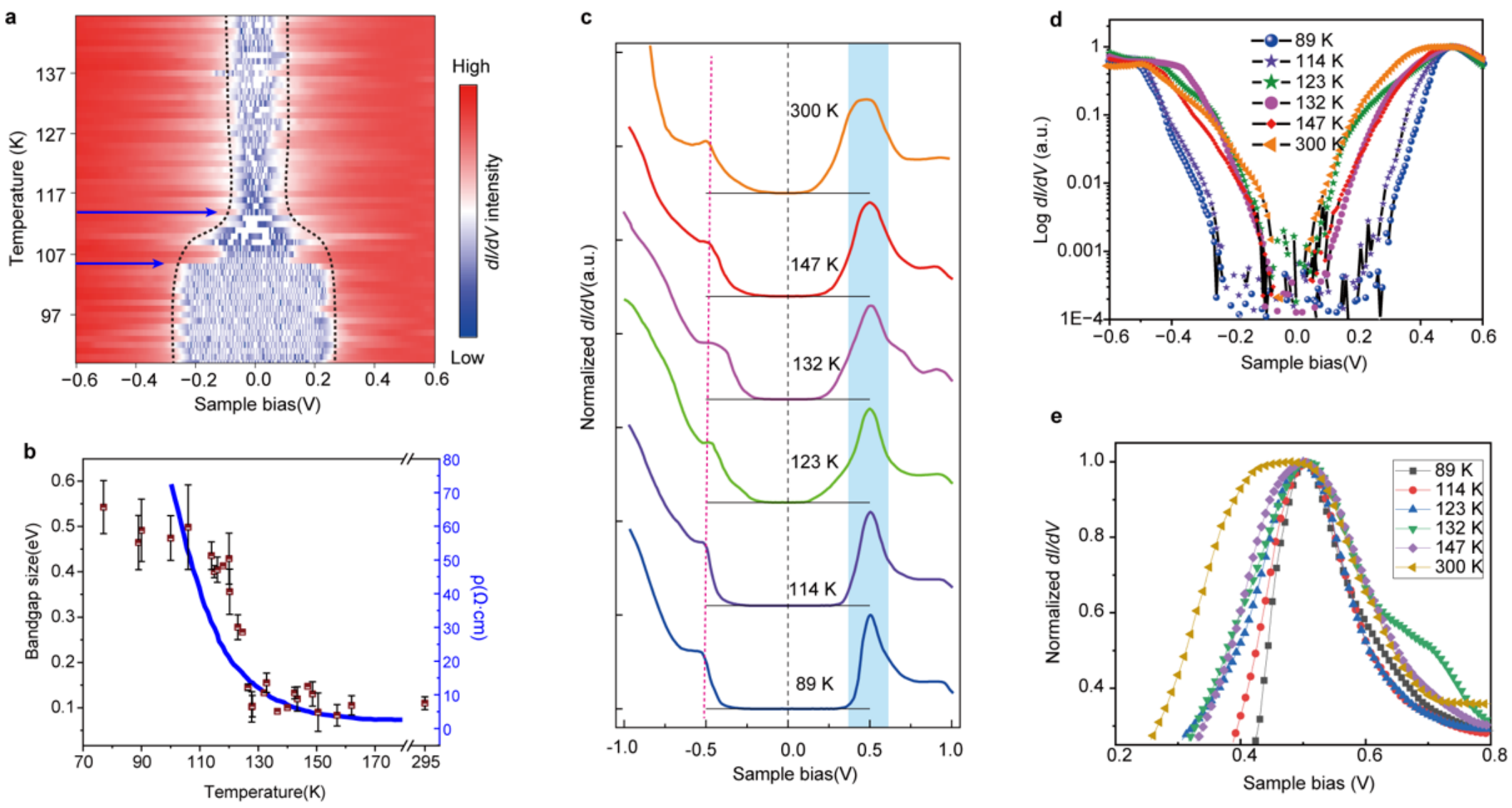

Fig. 3. Temperature-dependent *dI/dV* spectra on monolayer α-$RuCl_3$. (a) *dI/dV* spectra as the function of temperatures taken on monolayer α-$RuCl_3$ ($V_{bias}$ = 0.5 V, $I_{set}$ = 300 pA); (b) Plot of the charge gap magnitude of monolayer α-$RuCl_3$ as a function of temperature. The observed gap softening correlates well with the rapid decrease in resistivity, as indicated by the solid line derived from electron transport measurements in ref. [58]; (c) A series of normalized *dI/dV* spectra taken on monolayer α-$RuCl_3$ during warming of the sample displays the evolution of the charge gap as a function of temperature; (d) The logarithm plot of the averaged *dI/dV* curves in (c); (e) Zoomed-in the spectra of the shadow regime in (c) show the evolution of the peaked DOS as variation of temperature.

## V. DISCUSSION

The foregoing observations clarify that both the bulk and monolayer samples exhibit rapid gap softening at the critical temperature where spin correlations vanish. In contrast, negligible change occurs in the *dI/dV* spectra as temperature increases across the transition from the zigzag-ordered antiferromagnet to the KQPM phase. This suggests that spin correlations, whether manifesting as long-range order (as in the antiferromagnetic states below 7 K) or short-range frustration (mediated by Kitaev coupling), alongside the Coulomb repulsion $U$, significantly contribute to the formation of Mott-Hubbard bands in α-$RuCl_3$. Actually, parallel phenomena have also been reported in the $A_2IrO_3$ (A = Na, Li) systems, another prominent family of Kitaev QSL candidates. Their insulating states have been found to be intimately linked to magnetic spin correlations, fueling ongoing debate regarding whether these materials function as Mott insulators or Slater insulators [63–67]. Such ambiguity can further extends to other 4*d* and 5*d* transition metal compounds exhibiting the SOC assisted $J_{eff}$=1/2 Mott insulating states, exemplified by $Sr_2IrO_4$ [68–70]. This accumulating evidence implies that the spin correlations play a crucial role in shaping the Mott band structure, potentially representing a common feature in a series of SOC-driven Mott insulators.

Recent years have also witnessed intensified theoretical focus on the role of spin correlations within the Mott-Hubbard framework and its associated Mott transitions. Notably, theoretical models have successfully described continuous metal-insulator transitions in the QSL states of Mott insulators [17–19]. These models demonstrate that spin-charge separation, coupled with emergent gauge fields driven by spin frustration, induces a divergence in the quasiparticle effective mass, giving rise to insulating states with critical resistivity enhancement [18,19]. Another kinds of theoretical works employing the slave-fermion approach further explores Mott transition [71,72]. In this model, electrons are decomposed into spinons, holons and doublons, with a fermionic auxiliary field introduced to decouple kinetic energy terms. Critically, when antiferromagnetic on-site spin correlations are incorporated, holon-doublon pairing facilitates charge localization and enhances effective mass at finite temperatures. This renormalization of the high-energy charge sector by low-energy spin correlation is regarded as a strong energetic leverage effect [71]. Numerical calculations further reveal that spin correlations generate a hard Mott gap substantially larger than that predicted by the original Coulomb repulsion $U$ [72,73]. The charge density wave (CDW), often observed in strongly correlated systems, can also induce temperature dependence in the energy gap. However, in α-$RuCl_3$, the presence of a large intrinsic Mott gap (exceeding 2 eV in bulk) and the absence of a well-defined Fermi surface capable of nesting render the formation of a conventional Peierls-type CDW highly improbable. It is noteworthy that an unconventional incommensurate charge supermodulation has been observed in few-layer α-$RuCl_3$ on graphite [38,55,74]. These studies explicitly rule out a CDW origin and instead suggest alternative explanations, such as quantum oscillations of Majorana fermions, eleron-hole crystallization as well as hidden dipole orders.

Building on insights from these theoretical models and the research progress, we propose the physical framework depicted in Fig. 4: Within both the Kitaev QSL and spin-ordered states below $T_k$, spin correlations enhance the effective mass of the charge carriers and promote their localization under the Mott-Hubbard paradigm, consequently establishing a large hard Mott gap. Conversely, upon transition to the CPM phase where spin correlations vanish, the dissolution of emergent auxiliary fields reduces effective mass and enables delocalization of correlated charges – despite unchanged Coulomb repulsion $U$. This delocalization process drives the observed Mott gap softening and enhances charge conductivity in transport experiments [58]. However, as reported in Ref. [61,62], these correlated charges still exhibit ultralow carrier mobility and anomalously high effective mass due to the strong correlation within the Mott-Hubbard picture.

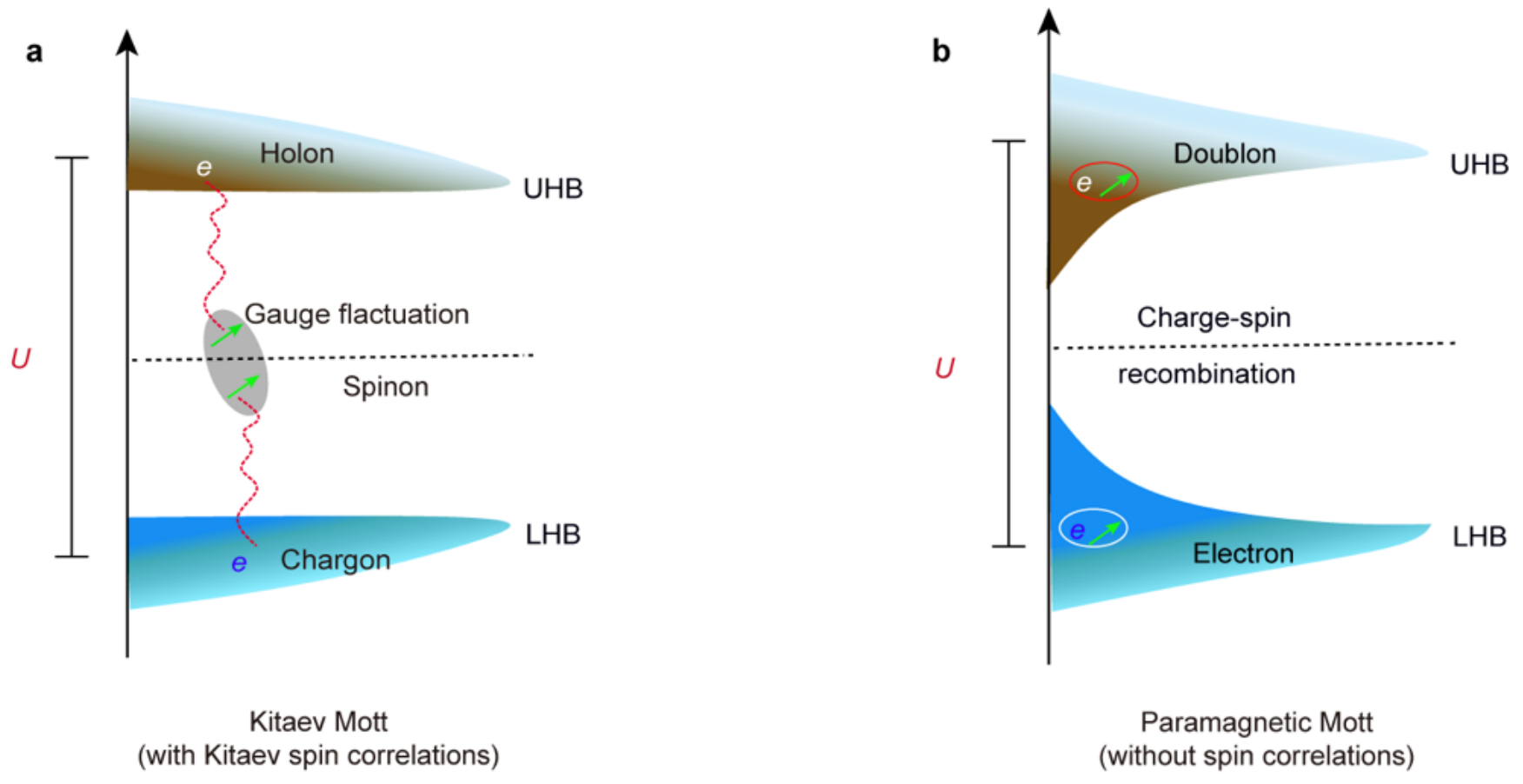

Fig. 4. Schematically illustrating the distinction between the KQPM Mott insulator and the CPM Mott insulator. (a) In the KQPM phase, while charges are gapped by Coulomb repulsion ($U$), emergent $Z_2$ gauge fluxes induced by spin correlations drive charge-spin separation. This process theoretically leads to divergence of the effective mass and enhances charge localization; (b) In CPM Mott states, recombination of chargons and spinons generates coherent in-gap states, delocalizing low-energy charge excitations within the Mott gap [18,19,72].

## V. CONCLUSION

In summary, we have investigated the temperature-dependent *dI/dV* spectra in both *in-situ* grown monolayer and exfoliated bulk α-$RuCl_3$. Both samples exhibit significant charge gap softening upon crossing the critical temperature regime from a Kitaev QSL Mott insulator to a CPM Mott insulator. By eliminating interference from structural phase transitions within the same temperature range, we establish that this Mott gap softening exhibits strong dependence on spin correlations. Specifically, spin correlations within the Mott insulator enhance the quasiparticle effective mass and promote charge carrier localization—a proposition consistent with several theoretical models. This mechanism successfully explains multiple previously unresolved experimental observations in α-$RuCl_3$. We further propose that a deeper understanding of the relationship between spin correlations and Mott gap softening within the Mott-Hubbard framework may provide insights into the pseudogap physics of high-temperature superconductors [71–73]. Our results also demonstrate that α-$RuCl_3$ provides an ideal platform for probing emergent Mott physics driven by multi-body correlations.

## ACKNOWLEDGMENTS

We would like to acknowledge helpful discussions with Prof. Tao Xiang. R. R. Du is funded by the Strategic Priority Research Program of Chinese Academy of Sciences (Grant No. XDB28000000), by the Innovation Program for Quantum Science and Technology (Grant No, 2021ZD0302600), and by the National Key Research and Development Program of China (Grant No. 2024YFA1409002). X. Z. is funded by Open Research Fund of State Key Laboratory of Materials for Integrated Circuits (SKLJC-K2025-05).